\UseRawInputEncoding
\documentclass[graybox]{svmult}
\usepackage{bookmark}
\usepackage{type1cm}
\usepackage{makeidx}
\usepackage{graphicx}
\usepackage{multicol}
\usepackage[bottom]{footmisc}
\usepackage{newtxtext}
\usepackage[varvw]{newtxmath}
\usepackage{natbib}
\usepackage{packages/journals}
\usepackage{hyperref}
\hypersetup{
    colorlinks=true,
    linkcolor=blue,
    citecolor=blue,
    filecolor=magenta,      
    urlcolor=blue,
}

\bibpunct{(}{)}{;}{a}{}{,}

\graphicspath{{./figures/}}

\makeindex

\newcommand*{\wasyfamily}{\fontencoding{U}\fontfamily{wasy}\selectfont}
\newcommand*{\jupiter}{{\text{\wasyfamily\char88}}}
\newcommand*{\saturn}{{\text{\wasyfamily\char89}}}

\begin{document}

\title*{Long-period transiting exoplanets: advances in detection and characterization}

\author{Sol\`{e}ne Ulmer-Moll\orcidID{0000-0003-2417-7006} and\\ Babatunde Akinsanmi\orcidID{0000-0001-6519-1598} and\\ Simon M\"uller\orcidID{0000-0002-8278-8377}}
\institute{Sol\`{e}ne Ulmer-Moll \at Leiden Observatory, Leiden University, Einsteinweg 55, 2333CA Leiden, The Netherlands, \email{ulmer-moll@strw.leidenuniv.nl}
\and Babatunde Akinsanmi \at Observatoire de Gen\`{e}ve, Universit\'{e} de Gen\`{e}ve. Chemin Pegasi 51, 1290 Versoix, Switzerland, \email{tunde.akinsanmi@unige.ch}
\and Simon M\"uller \at Department of Astrophysics, University of Z\"urich, Winterthurerstrasse 190, 8057 Z\"urich, Switzerland,
\email{simonandres.mueller@uzh.ch}}

\maketitle


\abstract{Most detected transiting planets have orbits which would fit within the one of Mercury, exposing them to intense stellar irradiation and interactions that significantly alter their properties. In contrast, colder planets with longer orbital periods are less affected, offering crucial insights into their formation and migration histories. Characterizing transiting warm and temperate planets is a key missing piece in the exoplanet puzzle. Dedicated photometric and spectroscopic follow-up of transiting events detected in space-based photometric data opened the way to detecting long-period transiting exoplanets. The wealth of information available for these transiting planets makes them golden targets for in-depth characterization. For giant planets, combining precise masses, radii, and ages with state-of-the-art planetary evolution models allows the estimation of their planetary bulk compositions, a crucial element to explore their formation and evolution pathways. Furthermore, these planets are compelling candidates for hosting moons and circumplanetary rings—features that could illuminate dynamical histories, satellite formation processes, and even potential habitable environments.}

\section{Introduction}
\label{sec:1}

Hot Jupiters, gas giants with orbital periods up to ten days, are easily observable in transit, while longer period planets remain difficult to detect because of rare transit events and lower transit probabilities. 
However, the properties of hot Jupiters are greatly influenced by their host star. Strong tidal interactions tidally lock planets and circularize their orbits (e.g. \citealt{eggleton_1998}). The high level of stellar irradiation induces atmospheric erosion and radius inflation mechanisms (e.g. \citealt{thorngren_2024}). In contrast, warm Jupiters, planets with orbital periods between 10 to 200 days or equilibrium temperature below 1000 K, are less influenced by their host star. Less prone to tidal realignment and stellar induced atmospheric changes, warm Jupiters are more likely to retain their original orbital and planetary properties post formation and migration processes.

The precise masses and radii of transiting warm planets coupled with planet evolution models enable the measurement of the planetary bulk metallicity and metal enrichment relative to the host star (e.g. \citealt{thorngren_2019e}). In addition, transit spectroscopy is a powerful technique for determining the atmospheric composition and metallicity (e.g. \citealt{madhusudhan_2019}). These bulk and atmospheric properties reflect the formation history of the planet with regard to its starting location, migration in the protoplanetary disk, and the nature of the accreted material. The exoplanet mass - metallicity relations constrain how gases and solids can accrete onto the growing planets (e.g. \citealt{hasegawa_2018}). Recent evidence suggests that the mass-metallicity relation may depend on the planetary orbital period \citep{dalba_2022}. Atmospheric metallicities can also be used to explore the exoplanet mass - metallicity (see review by  \citealt{swain_2024}).

If not formed in situ, warm Jupiters are thought to start their growth beyond the ice line and subsequently migrate inward to be found at their current location (e.g. \citealt{lin_1996}). 
Two main migration pathways have been put forward to explain the properties of the warm Jupiter population : disk-driven migration (e.g. \citealt{baruteau_2014}) and high-eccentricity migration (e.g. \citealt{rasio_1996}). The growing number of transiting warm giants provides opportunities for measuring the projected spin-orbit angle, a crucial tracer of the migration history of close-in planets. Constraints from Rossiter-McLaughlin observations suggest that warm Jupiters in single star systems are more aligned than hot Jupiters, hinting at distinct formation channels between the two populations (e.g. \citealt{rice_2022,wang_2024}).

%
In this chapter, we report the effort to detect new transiting warm giant planets orbiting bright stars. We review recent results regarding the characterization of warm giants in terms of interior structure modeling. Finally, we discuss the prospects for the detection of moons and rings around these longer period planets.

\section{Detection of the long-period transiting planets}
\label{sec:12}

\subsection{Photometric surveys}

Ground-based photometric surveys such as SuperWASP \citep{pollacco_2006a}, HATNet \citep{bakos_2007}, and KELT \citep{pepper_2007}, carried out blind searches and discovered a large number of transiting hot Jupiters with orbital periods of a few days. Longer period exoplanets are notoriously difficult to detect in transit. The transit probability is inversely proportional to the semi-major-axis, making the transits of long-period planets rare. Space-based photometric surveys such as CoRoT \citep{auvergne_2009}, Kepler \citep{borucki_2010}, and K2 opened the door to the discovery of some of the first transiting exoplanets with orbital periods from a few tens to hundreds of days (e.g. CoRoT-9b, \citealt{deeg_2010}; Kepler-420 Ab, \citealt{santerne_2014}; Kepler-1514 b, \citealt{dalba_2021a}). With a 4-year baseline, the Kepler mission was ideal for detecting transiting long-period planets. Unfortunately, a lot of these discoveries were made around relatively faint host stars, 
hindering the radial velocity follow-up and mass determination of these objects.

The Transiting Exoplanet Survey Satellite (TESS; \citealt{ricker_2015b}) is a space-based photometric mission that started scientific observations in July 2018. TESS aims at detecting transiting exoplanets orbiting bright host stars. TESS performs a near all-sky survey, each field is observed for 27 days before the telescope switches to another field. At high ecliptic latitudes, the TESS sectors overlap leading to nearly continuous observations for one year and allowing for consecutive transits of long-period planets to be detected. At the end of the second mission extension (2023-2025), TESS will have observed more than 80\% of the sky at least twice. Planets with an orbital period longer than the survey's baseline usually appear as single transit events in one TESS sector. We refer to these detections as monotransit candidates. Repeated observations of the same field by several TESS sectors give us access to longer period planets, and monotransit candidates can show a second and non-consecutive transit in a subsequent TESS sector, promoting them to duotransit candidates. 

The detection of monotransit candidates is expected to yield a large number of long-period planets (e.g. \citealt{cooke_2018b,cooke_2019a}). \citet{villanueva_2019b} predicted that close to 1100 single transit events can be detected in the TESS data during the primary mission (Years 1 and 2, observations of the southern and northern ecliptic hemispheres). Using the actual TESS window functions and noise properties, \citet{rodel_2024a} estimated that about 400 exoplanets with orbital periods larger than 25 days and 110 exoplanets with orbital periods larger than 100 days should be detectable from the observations of the southern ecliptic hemisphere in Years 1 and 3. 

\subsection{Identification and vetting of transiting candidates}

\subparagraph{Single transit event detection} 
Dedicated searches for monotransit and duotransit candidates are performed with various techniques that do not rely on the periodicity of the transit signal, but focus on detecting a single transit event while usually accounting for stellar and/or instrumental effects. Classical methods based on least squares minimization have been implemented to fit transit models with different transit durations, for example, in \citet{osborn_2016a} and \citet{gill_2020c}. \citet{hawthorn_2024} search for consecutive photometric points considered as outliers compared to the median of the light curve to detect monotransit events. Recently, neural networks, trained on Kepler data \citep{cui_2021,hansen_2024}, TESS data \citep{salinas_2025}, and PLATO simulated light curves \citep{vivien_2025}, have also been implemented with the aim of discovering single transit events. These searches led to the announcement of tens of new candidates from TESS data alone, e.g., 85 duotransits from Year 1 and 3 \citep{hawthorn_2024} and 88 monotransits from Year 1 and 2 \citep{salinas_2025} Additionally, the Science Processing Operations Center pipeline (SPOC, Jenkins et al. 2016), the Quick-look pipeline (QLP, \citealt{huang_2020,huang_2020a}), and algorithms based on Box least square fitting (e.g. \citealt{montalto_2023}) provide transiting candidates detected from single transit events even though these pipelines are optimized to identify repeated transit events. The citizen science project, Planet Hunters TESS, searched the SPOC light curves and reported 73 monotransit candidates from TESS Year 1 and 2 observations \citep{eisner_2021}.

\subparagraph{Orbital period estimation} 
In the case of monotransits, the orbital period of the object is unknown apriori, and the modeling of the transit requires a different fitting approach. The orbital period of monotransits can be estimated with Equation 1 from \citet{wang_2015b} based on Equations 18 and 19 from \citet{winn_2010}: 
\begin{equation}
\frac{P}{1\,\rm yr} = \left(\frac{T}{13\,\rm hr} \right )^3.\left(\frac{\rho}{\rho_\odot}\right).\left(1-b^2\right)^{-\frac{3}{2}}
\end{equation}
where $P$ is the orbital period, $T$ is the transit duration, $\rho$ is the stellar density, $\rho_\odot$ is the solar density, and $b$ is the impact parameter. Constraining the stellar density is essential to obtain precise constraints on the orbital of the monotransit candidates. \citet{sandford_2019} combined stellar radius measurements from Gaia DR2 parallaxes and stellar masses from isochrone fitting to determine the orbital period of K2 long-period planets. \citet{magliano_2024} investigated the effect of eccentricity and argument periastron on the analytical derivation of the orbital period and inclination of monotransits. For planets first discovered as monotransits, they show how their orbital period estimates compare to the published solutions based on additional transit and radial velocity observations. \citet{osborn_2022e} developed \texttt{MonoTools}, a software package that is tailored to analyze systems hosting monotransits and duotransits. The transit duration, impact parameter, and radius ratio combined with stellar density are fitted to constrain the transverse planetary velocity, which can be converted to the orbital period. In the case of duotransits, only a given number of orbital period aliases are allowed by the photometric data. \texttt{MonoTools} marginalizes over the valid period aliases and computes the posterior probabilities for each period alias. In the future, PLATO's asteroseismology program will bring stronger constraints on the stellar parameters, thanks to the continuous observation during 2 years of the PLATO LOPS2 field and the short photometric cadence of 25 seconds or even 2.5 seconds for the fast cameras \citep{rauer_2025}. The stellar density measurements for a sample of bright dwarf and sub-giants stars will help precisely estimate the orbital period of monotransits.

\subparagraph{Photometric and spectroscopic vetting} 

Several instrumental or astrophysical effects can produce a transit-like signal in the light curve. The vetting of the detected transits is an essential first step before investing further resources in the follow-up of such signals. Asteroid crossings, blended or on-target eclipsing binaries, systematic effects, and stellar variability can mimic a transit signal, but these effects can also be excluded thanks to several diagnostic tools such as the background flux, the x and y centroid positions, or the individual pixel light curves around the target star. The vetting and classification of the TESS transiting candidates based on these diagnostic tools have been implemented in software packages such as \texttt{LATTE} \citep{eisner_2020a} and \texttt{TRICERATOPS} \citep{giacalone_2020}. Machine learning algorithms such as random forests (e.g. \citealt{montalto_2020b}) and convolutional neural networks (e.g. \texttt{NotPlaNET}, \citealt{tardugnopoleo_2024}) are also used to classify transit events. For more details, \citet{collins_2018} present the vetting and follow-up processes developed in the context of the KELT survey. These methods are applied for the detection and confirmation of TESS planet discoveries. \citet{magliano_2024} show examples of false positive scenarios in the case of monotransit detections.

Spectrocopic vetting is critical to confirm that the transiting candidates are in the planetary regime and not low-mass eclipsing binaries or brown dwarfs. Reconnaissance spectroscopy is mostly performed with high-resolution spectrographs on 1 to 2-m class telescopes. Two spectra taken a few days to weeks apart are sufficient to rule out single-line and double-line spectroscopic eclipsing binaries. These first spectroscopic observations allow one to extract the stellar parameters and confirm that the star is appropriate for precise radial velocity observations. Giant stars are usually excluded because of the difficulty in interpreting the observed radial velocity variations (e.g., \citealt{delgadomena_2023}). Rapidly rotating stars are not conducive to high-precision radial velocities as the rotation broadens the stellar lines, leading to poor radial velocity precision (e.g. \citealt{bouchy_2001a}).
Photometric and spectroscopic vetting allows to curate a list of suitable transiting candidates before investing more observing time with highly demanded high-resolution spectrographs or other photometric facilities.


\begin{figure}
    \centering
    \includegraphics[width=0.9\linewidth]{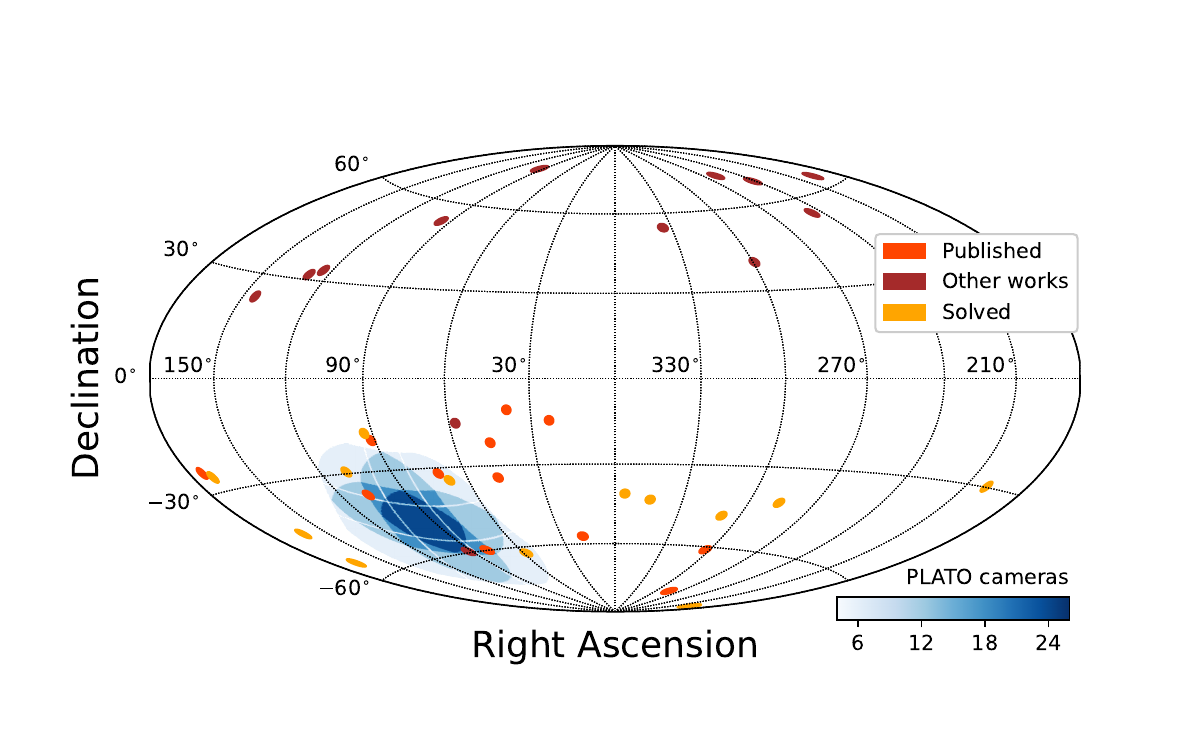}
    \caption{All sky Aitoff projection of the TESS transiting giant planets with orbital periods ranging from 20 to 260 days. Published (and solved) planets led or co-authored by PlanetS researchers are represented with orange dots (yellow dots). Planets published by other teams (brown dots). First PLATO field is shown in blue.}
    \label{fig:map_monos}
\end{figure}

\subsection{Confirmation of long-period transiting planets}


Photometric follow-up observations are required to confirm the orbital period of the long-period transiting candidates detected in the TESS data. Simulations from \citet{cooke_2021} show that duotransits from TESS observations in Year 1 and 3 have an average of 38 possible period aliases. The Next Generation Transit Survey (NGTS, \citealt{wheatley_2018}) has a multi-year campaign to observe mono and duotransit candidates. Monotransit candidates are followed in a blind survey mode, where the star is observed every possible night until a transit event matching the TESS transit is detected. Duotransit candidates are scheduled for observations when one of the period aliases is predicted to transit. Archival NGTS photometry is also used to recover transit events prior to TESS transit detection (e.g., \citealt{battley_2024,gill_2024}). The Swiss-led ESA mission CHEOPS \citep{benz_2021,fortier_2024} is dedicated to the characterization of exoplanets and the CHEOPS Guaranteed Time Observers (GTO) has an ongoing program to confirm the orbital period of small long-period planets seen as duotransits (e.g. \citealt{osborn_2022e}). Duotransit candidates with larger transit depths, between 2 to 4 ppt, are scheduled for short filler observations, as CHEOPS is able to successfully recover their transit with only partial transit observations (e.g. \citealt{ulmer-moll_2023}). 
Other ground and space-based photometric facilities such as ASTEP \citep{mekarnia_2016}, LCO \citep{brown_2013}, and NEOSSat \citep{laurin_2008} observe duotransit and long-period transiting candidates (e.g. \citealt{hobson_2023}). Once the orbital period of a transiting planet is known, photometric observations are still valuable to prevent ephemeris deterioration as highlighted by \cite{dragomir_2020}.

Radial velocity follow-up is necessary to confirm transiting candidates and establish their orbital periods, planetary masses, and eccentricities. Long-term radial velocity campaigns with stable high-resolution spectrographs are instrumental in confirming these long-period planets. We note interesting detections of giant planets with orbital periods larger than 100 days (e.g. TOI-2180 b, \citealt{dalba_2022}; TOI-2010 b, \citealt{mann_2023}; TOI-2449 b, Ulmer-Moll, under review) which are ideal systems to search for exomoons \citep{szabo_2023}. The population of warm Jupiters displays a wide range of eccentricities, which can be explained by different migration pathways (e.g., \citealt{dawson_2018}) or planet-planet scatterings (e.g. \citealt{carrera_2019}). New TESS detections confirm this diversity, notably with the characterization of several warm Jupiters on highly eccentric orbits (e $>$ 0.5) : e.g. TOI-2338 b and TOI-2589 b \citep{brahm_2023}, TOI-2134 c \citep{rescigno_2024}, TOI-5110 b \citep{heidari_2025}. The detection of a massive warm Jupiter with an orbital eccentricity of 0.94 by \citet{gupta_2024} suggests that this planet is seen in the middle of inward migration to become a hot Jupiter.
In the last five years, the discoveries of transiting warm giant planets with precise masses and radii (20\% and 8\% uncertainties, respectively) around bright stars doubled, demonstrating the success of the photometric and radial velocity follow-up efforts. For planets with orbital periods larger than 20 days, 37 new transiting warm giant planets from TESS have been published, see Fig~\ref{fig:map_monos}. The work carried out in the context of the Monotransit Initiative from PlanetS led to the detection of 13 published planets, and 13 additional planets have been solved and will be published.




\section{Characterization of warm giant exoplanets}
\label{sec:2}

The characterization of giant exoplanets is essential for constraining theories of giant planet formation and evolution, and for placing the giant planets of the Solar System in a broader astrophysical context \citep{2015ApJ...810..105M,2020PASP..132j2001H,2021JGRE..12606643K}. Mass-radius measurements of moderately irradiated warm Jupiters are commonly employed to estimate planetary bulk composition - an important parameter for informing models of formation, internal structure, and long-term evolution. In this section, we examine the role of evolution models in inferring the bulk composition of warm giant planets, discuss major sources of theoretical uncertainty, and present a few characterization highlights from within the PlanetS network.

The majority of detected giant exoplanets are categorized as hot Jupiters; however, a subset comprises warm giants, which exhibit equilibrium temperatures below approximately 1000 K. Warm giants are of particular interest for characterization studies, as their physical properties are less influenced by the inflation mechanisms that affect hot Jupiters. The inflation observed in hot Jupiters is attributed to mechanisms that are not yet fully understood, complicating the modeling of their internal structures \citep[see, e.g.,][for a review]{2021JGRE..12606629F}.

\begin{figure}
    \centering
    \includegraphics[width=\linewidth]{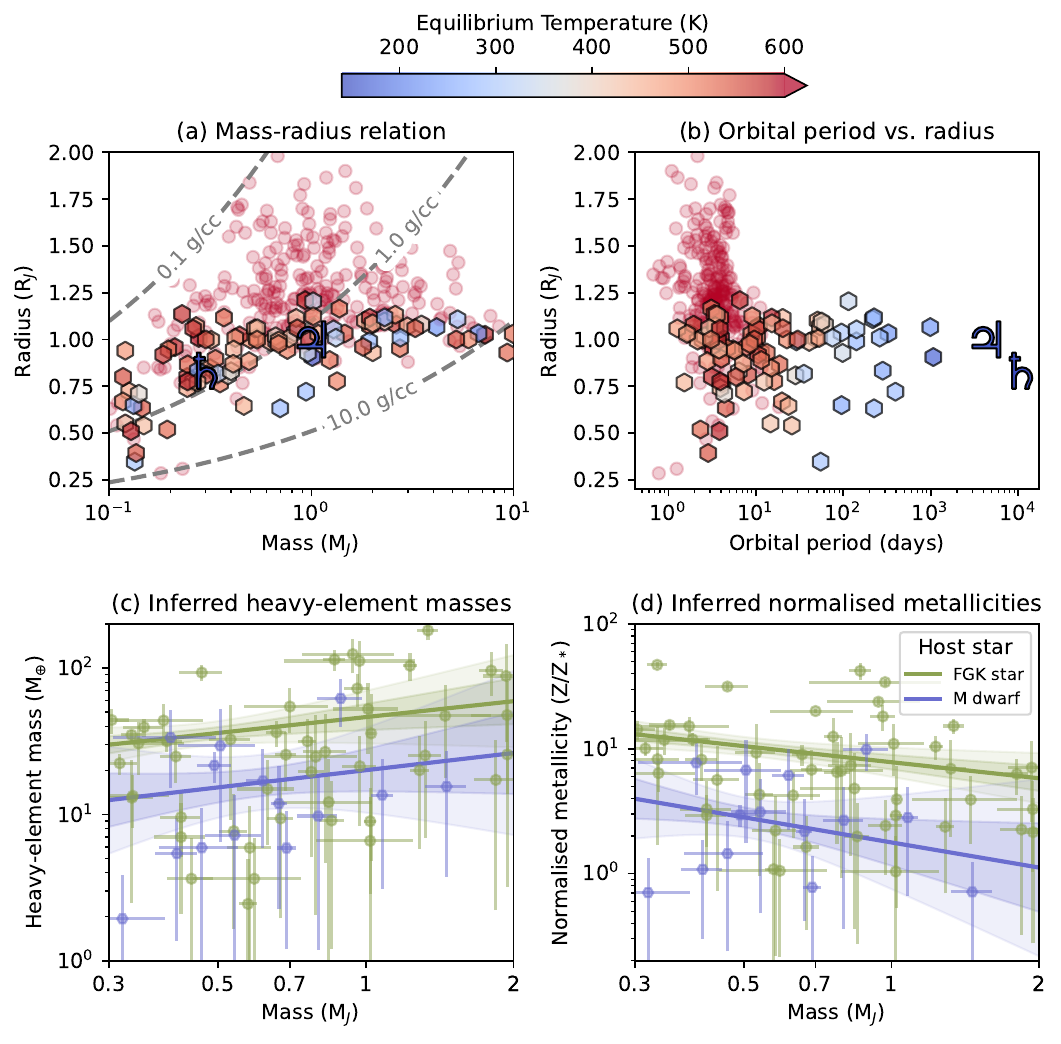}
    \caption{Mass-radius relation (a) and orbital period vs. radius (b) of observed giant exoplanets with masses between 0.1 and 10 M$_J$. The data is coloured according to the equilibrium temperature. Hexagons show the warm giant exoplanets with $T_{\textrm{eq}} < 1000$ K, and hot Jupiters are shown as small circles in the background. The mass-radius diagram shows the 0.1, 1 and 10 g/cc isochores for reference. For a comparison, Jupiter (\jupiter) and Saturn (\saturn) are drawn with their respective symbols. Panels (c) and (d) show the inferred relation between the planet mass (between 0.3 and 2 M$_J$) and their heavy-element masses and metallicities. The symbols and colours indicate the host-star type: Green circles for FGK stars and blue hexagons for M dwarfs. The lines and shaded regions show the best fit and the 1 to 2 $\sigma$ contours. The data were taken from the \textit{PlanetS catalog} \citep{parc_super-earths_2024} and \citet{2025A&A...693L...4M}.}
    \label{fig:warm_giant_exoplanets}
\end{figure}

In Figures \hyperref[fig:warm_giant_exoplanets]{2a} and \hyperref[fig:warm_giant_exoplanets]{2b}, we show the mass-radius relation and orbital period vs. radius of observed giant exoplanets from the \textit{PlanetS catalog} \citep{parc_super-earths_2024}. These data are based on reliable and robust mass-radius measurements of transiting planets. Despite recent advances in detecting long-period giant planets (see Section \ref{sec:12}), there is no detected giant transiting exoplanet with a similar orbital period to Jupiter and Saturn. At a given mass, there is a large variety of observed radii due to different incident stellar fluxes, planetary ages and, in particular, compositions. This is demonstrated in Figures \hyperref[fig:warm_giant_exoplanets]{2c} and \hyperref[fig:warm_giant_exoplanets]{2d}, where the inferred heavy-element masses and metallicities of giant exoplanets from \citet{2025A&A...693L...4M} are shown.

The substantial radii observed in giant exoplanets signify the presence of extensive hydrogen-helium envelopes, similar to Jupiter and Saturn, which undergo gradual contraction and cooling over time as the planets evolve \citep[e.g.,][]{1977Icar...30..305H,2001RvMP...73..719B}. Consequently, the characterization of these planets necessitates reliance on theoretical models that simulate their evolutionary processes. Combining planetary evolution models with measurements of a planet’s radius, mass, and the age of its host star allows for estimates of the planet’s bulk composition. In the case of giant planets, this composition is typically characterized by the total mass of heavy elements within the planet \citep[e.g.][]{2006A&A...453L..21G,2007ApJ...659.1661F,2016ApJ...831...64T}. These estimates are obtained by comparing the observed planetary radius to the radius predicted by theoretical evolution models.

To account for uncertainties in observational data (mass, radius and stellar age), the bulk metallicity is often inferred as a posterior probability distribution. Since giant planets undergo cooling and contraction over time, their evolutionary stage directly influences their observable properties. At a given planetary age, differences in radius translate into different inferred compositions, making accurate characterization dependent on robust measurements. Reliable determination of the heavy-element content in giant planets therefore requires an interplay between precise measurements of planetary radius, mass, age, and theoretical evolution models.

\subsection{Giant planet evolution models}

Giant planet evolution models are typically constructed assuming spherical symmetry and hydrostatic equilibrium. The evolution is governed by a set of coupled partial differential equations that describe the conservation of mass, momentum, and energy, as well as the transport of heat and chemical elements \citep[e.g.,][]{kippenhahn_stellar_2012}. In most models, energy transport is assumed to occur via radiation (and, to a lesser extent, conduction) or convection.

To determine the dominant mode of energy transport (convection, radiation or conduction) at a given location within the planet, the Schwarzschild \citep{schwarzschild1958structure} or the Ledoux criteria \citep{1947ApJ...105..305L} are commonly applied. The Ledoux criterion is more general and can be applied to planets with inhomogeneous compositions. When convection is present, more complex models use the mixing-length theory \citep{bohm-vitense_uber_1958} to treat convection as a diffusive process. In fact, convection is often efficient enough that the deep interiors of giant planets are nearly adiabatic \citep{2004jpsm.book...35G}, and simpler models skip using the mixing-length theory by assuming a perfectly adiabatic interior. To complete the model, additional inputs are required, including a suitable equation of state, a prescription for opacity, and appropriate atmospheric boundary conditions.

Because the evolution equations do not have general analytical solutions, they must be solved numerically \citep[e.g.,][]{1965ApJ...142..841H, 1995A&AS..109..109G, 1997ApJ...491..856B, 2015ApJ...803...32V,2024A&A...688A..60A,sur_apple_2024}. In the PlanetS network, three codes are used to model the evolution of giant planets: \texttt{Completo21} \citep{2012A&A...547A.111M, jin_planetary_2014, 2019A&A...623A..85L, 2024A&A...692A.202P}, \texttt{planetsynth} \citep{2021MNRAS.507.2094M}, and a modified version \citep{2020A&A...638A.121M,2020ApJ...903..147M, muller_can_2024} of the stellar evolution code Modules for Experiments in Stellar Astrophysics (MESA; \citet{2011ApJS..192....3P,2013ApJS..208....4P,2018ApJS..234...34P,2019ApJS..243...10P,2023ApJS..265...15J}). Discussing the details of these codes is beyond the scope of this review, and we refer the reader to the respective articles.

\subsection{Uncertainties in giant planet evolution models}

Models of giant planet evolution rely on a variety of assumptions, including the equations of state for constituent elements, their spatial distribution within the interior, the choice of representative materials for heavy elements, initial conditions, energy transport mechanisms, atmospheric structure, and opacity. These parameters significantly influence the planet’s thermal evolution, thereby affecting key model predictions such as radius and luminosity \citep{1999Sci...286...72G}. Here, we briefly introduce the most influential modeling uncertainties, and note that a more thorough discussion was presented in \citet{2023FrASS..1079000M}.

Several studies have shown that the hydrogen-helium equation of state used in evolution models strongly affects the predicted radii \citep{2020ApJ...903..147M,2023A&A...672L...1H, 2025A&A...693L...7H}. This is largely due to updated hydrogen equations of state \citep{2019ApJ...872...51C}, and a better understanding of hydrogen-helium mixtures \citep{2021ApJ...917....4C,2023A&A...672L...1H}. Furthermore, the details of the outer boundary conditions of the evolution models strongly influence the cooling. Key uncertainties include the radiative opacities and whether there are grains or clouds present \citep{2013MNRAS.434.3283V,2014A&A...572A.118M, 2019Atmos..10..664P}, and atmospheric models properly accounting for the stellar irradiation \citep{2007ApJ...659.1661F,2010A&A...520A..27G, 2014A&A...562A.133P}. Finally, there is also the complication of unknown formation scenarios and initial conditions, namely whether giant planets have hot, warm or cold starts \citep[e.g.,][]{2003A&A...402..701B,2007ApJ...655..541M,2012ApJ...745..174S,2017ApJ...846L..17B}. The initial conditions influence the dominant modes of energy transportation and the distribution of heavy elements in the interior, which then affect the evolution \citep[e.g.][]{2008A&A...482..315B,2012A&A...540A..20L,2013MNRAS.434.3283V,2020A&A...638A.121M,2025ApJ...979..243A,2025arXiv250412118K}.

These uncertainties have direct consequences for the characterization of giant exoplanets: Currently, the theoretical uncertainties in estimated bulk compositions of warm giant exoplanets exceed those that come from uncertainties in observed parameters (planetary mass, radius and age). Therefore, the topics mentioned above are key areas of research that should be addressed in future research.

\subsection{Connection between planetary interiors and atmospheres}

The James Webb Space Telescope (JWST; \citet{gardner_james_2006}) and the upcoming ARIEL mission \citep{2018ExA....46..135T} promise to usher in a new era of exoplanet characterization with their atmospheric measurements of many giant exoplanets. Determining the chemical composition of giant exoplanetary atmospheres represents a transformative step in the characterization of these bodies. Atmospheric abundances offer critical insights into the planets' internal structures and formation histories \citep[e.g.][]{2014PNAS..11112601B,2019AJ....158..239T,2021ApJ...909...40T, 2022AJ....164...15E}. Although establishing a definitive relationship between atmospheric and bulk compositions remains a complex challenge \citep[e.g.,][]{2022ExA....53..323H}, atmospheric measurements of warm Jupiters provide an opportunity for improved characterization and constraining formation scenarios. Notably, atmospheric measurements have been shown to resolve degeneracies in bulk composition estimates, potentially reducing the associated uncertainties by a factor of four \citep{2023A&A...669A..24M}.

Despite these advancements, a fundamental question remains unresolved: What is the precise relationship between atmospheric and interior compositions? Observations of Solar System gas giants suggest that atmospheric and internal structures can differ significantly \citep[e.g.,][]{2024arXiv240705853H}. Systematic atmospheric characterizations of a large sample of giant exoplanets will enable more robust constraints on interior models and provide insights into the degree of compositional mixing \citep{2019ApJ...874L..31T}. It is important to note that atmospheric measurements probe only the uppermost atmospheric layers, and the extent to which these reflect deeper compositional gradients is uncertain. Additionally, given the low mass and radiative nature of these outer layers, observed compositions may be influenced by recent exogenous enrichment events, such as the accretion of small bodies that undergo atmospheric ablation \citep{howard_exploring_2023,muller_can_2024}. Therefore, understanding the conditions under which atmospheric composition reliably reflects bulk properties remains a critical objective for exoplanetary science.

\subsection{Recent advances in the characterization of warm giant exoplanets}

In this section, we report on a few highlights of exoplanet characterization within the past few years and focus on studies carried out in the PlanetS network.

The updated \textit{PlanetS catalog} provides a robust list of planets with measured masses and radii. These data recently yielded updated mass-radius relations of low- to high-mass exoplanets \citep{2024A&A...686A.296M, parc_super-earths_2024}. Regarding giant planets, \citet{2024A&A...686A.296M} found that the transition from intermediate-mass to massive gaseous planets occurs at a mass of about $130 M_\oplus$. This is consistent with the idea that Saturn could be imagined as a "failed" gas giant with its mass close to the transition region \citep{helled_mass_2023}.

As mentioned in Section \ref{sec:12}, there were several detections and characterizations of long-period transiting giant planets \citep[e.g.,][]{2022A&A...666A..46U,2023AJ....165..227B,2024A&A...686A.230B,2025A&A...693A.144G}. As discussed previously, these planets are particularly interesting for characterization since their properties are less affected by the stellar irradiation, and they can be more readily compared to the solar system gas giants. In recent years, there were also more discoveries of both very high- and low-density warm giant planets \citep[e.g.,][]{2023A&A...672L...7K,2024NatAs...8..909B}, as well as more Saturn and sub-Saturn warm giant planets \citep[e.g.,][]{2023A&A...675A..39P,2024MNRAS.532.1612H}. In particular, \citet{2024MNRAS.532.1612H} presented two high-density sub-Saturns ($M_p \simeq 0.2 M_J$). Their characterization revealed metal-rich interiors, challenging current formation theories, which generally predict lower metal enrichment for similar planets.

Over the past few years, there were several detections and characterizations of giant planets with M-dwarf host stars \citep[e.g.,][]{2023AJ....166...30C,2024ApJ...962L..22D,2024AJ....167....4H,2024AJ....168..235K,2024AJ....168..273B,2025AJ....169..187R}. These planets are exciting for many reasons \citep[e.g.,][]{2024AJ....167..161K}: M-dwarf stars are the most common stars in the galaxy \citep[e.g.,][]{2021A&A...650A.201R}, and there are therefore many potential targets. Due to the low effective temperatures of M dwarfs, even planets in close orbits are within the warm Jupiter regime. Finally, the discovery of giant planets around M dwarfs challenges formation theory, which predicts that they are unlikely to form due to lower protoplanetary disk masses and less available solids \citep[e.g.,][]{2008ApJ...673..502K,andrews_mass_2013,pascucci_steeper_2016,schlecker_rv-detected_2022}. Recently, \citet{2025A&A...693L...4M} investigated the bulk metallicity of giant planets around FGK and M-dwarf stars. As shown in Figures \hyperref[fig:warm_giant_exoplanets]{2c} and \hyperref[fig:warm_giant_exoplanets]{2d}, they found a lack of metal-rich planets around M dwarfs (compared to FGK stars). This could imply different formation conditions or pathways for these two populations, however, a robust conclusion will require further measurements of giant planets around M dwarfs. As discussed previously, atmospheric measurements are another exciting avenue to improve the characterization of giant exoplanets. The PlanetS network is involved in a JWST observation program to characterize the atmospheres of seven giant planets around M dwarfs \citep{2023jwst.prop.3171K}. The first results for TOI-5205b were recently reported in \citet{2025arXiv250206966C}. The planet appears to have a sub-solar atmospheric metallicity but a metal-rich interior, suggesting a decoupling between the planetary interior and atmosphere, such that the planet is not well mixed.

\section{Moons and Rings around Long period planets}
\label{sec:3}
Moons and rings are ubiquitous around the planets of the solar system; the giant planets particularly host several tens of moons and are adorned by complex ring systems. The search for these natural satellites around exoplanets—referred to as exomoons and exorings—has become an exciting and rapidly developing area of exoplanetary research \citep{Heller2018, Teachey2018HEK.I}. These features hold profound implications for understanding planetary system formation, dynamics, and even the potential for habitability. Long-period exoplanets are particularly promising targets to search for these features. Similar to the giant planets of the solar system, their wide orbits (with periods of several tens to hundreds of days) offer a dynamically stable environment where moons and extensive ring systems can form and survive over long timescales. In this section, we examine the exciting possibility of detecting moons and rings around long-period exoplanets. The section discusses the scientific motivation and implication of detecting these features, summarizes the major detection techniques, the outcome of searches, and continued efforts within PlanetS to detect these features.

\subsection{Motivation}
Moons and rings encode the history of planetary systems, and detecting them around exoplanets can offer new insights to better understand the complex architectures and diversity of planetary systems. Moons, for instance, may form in situ within the circumplanetary disk from giant impacts (as with Earth's moon) or from gravitational capture of unbound objects. Their orbits, compositions, and dynamical interaction with the host planet provide insights into planetary migration, tidal forces, and the stability of systems. They can also extend habitability beyond the canonical planetary boundaries, as moons of long-period gas giants could maintain stable surface/subsurface conditions due to tidal heating and magnetic shielding from their host planets \citep{Heller2013HotStars}. On the other hand, rings may form around a planet from leftover circumplanetary material during planet/moon formation, or from the tidal disruption or collision of minor bodies/moons, or from ejected materials from an orbiting moon. Rings could reveal ongoing processes of moon formation via the presence of gaps, moon geophysical activity that feeds the ring system (e.g., Enceladus replenishing Saturn's E-ring), or resonant interactions that shape the orbits of ring particles and moons.

Ultimately, finding these satellites in extrasolar systems will provide a more statistically robust picture of how common these features are and how they depend on the age and architecture of the system, the mass of the planet, the orbital distance, and the formation environment \citep{Canup2002FormationAccretion}.

\subparagraph{Why long-period planets?}
Generally, a moon orbits at some distance between the planet's Roche limit ($R_{\mathrm{Roche}}$; below which it is broken apart by tides) and its Hill radius ($R_{\mathrm{Hill}}$; beyond which it is no longer bound to the planet). More specifically, \citet{Domingos2006StablePlanets} showed that prograde moons are only stable within 0.49\,$R_{\mathrm{Hill}}$ and retrograde ones further out to 0.93\,$R_{\mathrm{Hill}}$. Since the size of a planet's Hill radius is proportional to its orbital distance to the star, planets with longer orbital periods are more favorable moon hosts. Conversely, rings typically form within $R_{\mathrm{Roche}}$, since particles outside this limit tend to coalesce to form moons. The long-term stability of rings around a planet requires that $R_{\mathrm{Roche}} < 2/3 R_{\mathrm{Hill}}$ \citep{vieira}. Therefore, hosting rings within the Roche limit of a planet requires the planet to have a large enough Hill radius, which might not be possible for short-period planets \citep{Kane2017moons}. Icy ring particles, as common in the solar system, can only be found beyond the snow line, whereas ring particles composed of denser rocky materials can survive around warmer planets.  As $R_{\mathrm{Roche}}$ varies inversely with the density of the ring particles, icy or low-density particles can form rings with large radial extents which increases their detectability, whereas denser ring particles around the shorter period planets will have to be tightly packed and more diffucult to discern.

\subsection{Methods for detecting exomoons and exorings}
Detecting moons and rings around exoplanets is quite challenging due to the subtle signals of these features, which still have to be disentangled from the noise and other astrophysical phenomena. Nevertheless, a variety of methods have been proposed and, in some cases, tentatively applied to search for their signatures. The methods discussed here exploit precise photometric and timing data from transiting exoplanets to infer the presence of either a moon or ring system around an exoplanet. 

\subparagraph{Detecting exomoons}
Transit Timing Variations (TTVs) and Transit Duration Variations (TDVs) are very strong observational signatures of exomoons that involve measuring small variations in the timing and duration of planetary transits \citep{Kipping2009a}. The gravitational influence of a moon can cause the planet to wobble around the barycenter of the system and also change its apparent velocity or transit chord \citep{Kipping2009b}. This induces predictable variations in the observed transit times (TTVs) and transit durations (TDVs). Plotting these variations as a function of transit epoch reveals that their sinusoidal signals are $\pi/2$ out of phase, with the TTV signal leading the TDV. The amplitudes of the signals depend on the moon’s mass, orbital distance, and the geometry of the system. When analyzed together, TTVs and TDVs can help distinguish moon-induced signals from those caused by additional planets or instrumental noise and also break the degeneracy of simultaneously measuring the mass and orbital distance of the moon \citep{Heller2016PredictableExomoons}. \cite{Kipping2020} and \cite{Kipping2020ImpossibleExomoon} further outline methods to distinguish TTVs due to exomoons from those from other astrophysical phenomena.

\begin{figure}[t]
    \centering
    \includegraphics[width=0.49\linewidth]{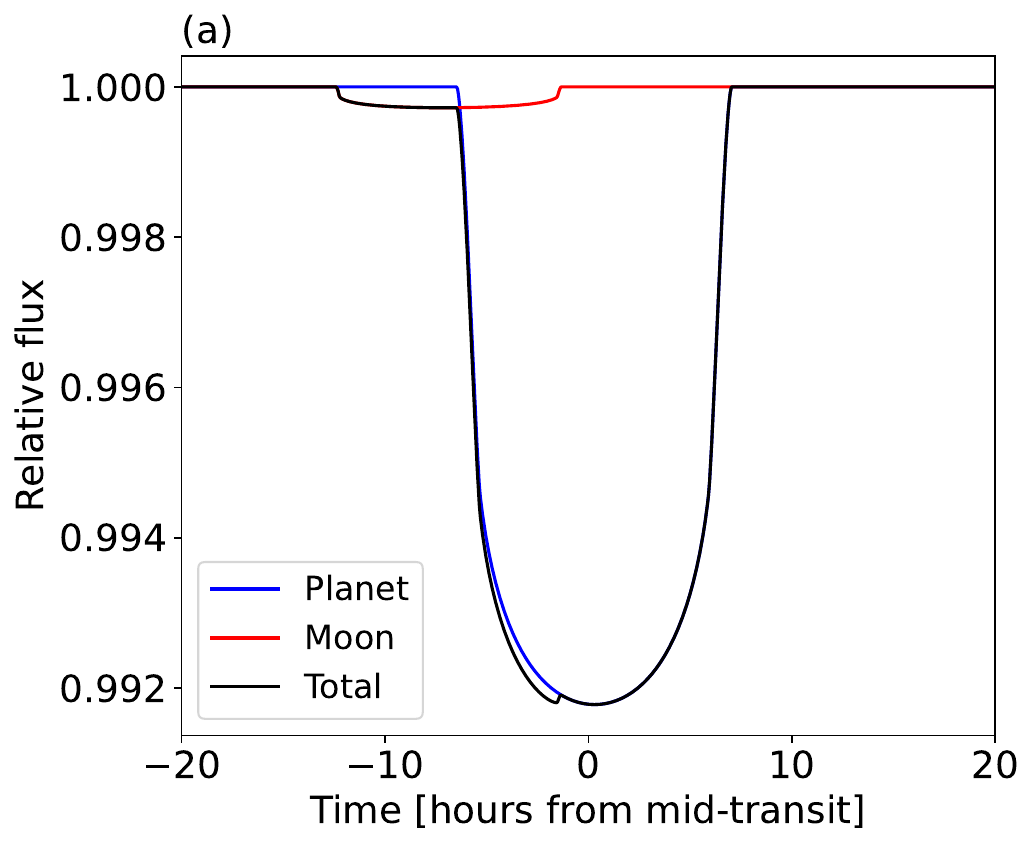}
    \includegraphics[width=0.49\linewidth]{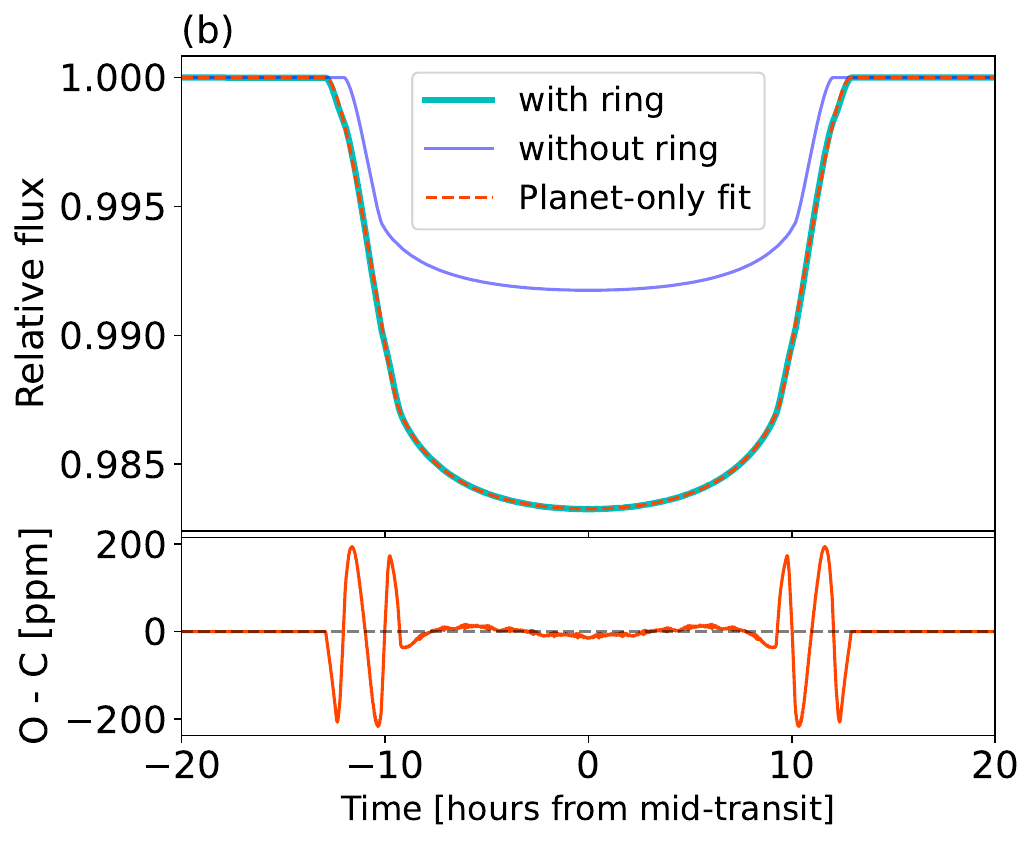}

    \caption{Sample transit light curves of moons and rings. (a) transit light curve of a planet with an orbiting moon where the moon is leading the planet so its transit (red) occurs before the ingress of the planet (blue). (b) transit light curve of a Saturn-like planet with rings (cyan) compared to the same planet without ring (blue). The planet-only fit to the ringed planet transit signal is shown in red and the residuals in the bottom panel. }
    \label{fig:moon_ring_transit}
\end{figure}

\bigskip
Similar to how planets cause dips in the stellar light as they transit, moons may also produce secondary, shallower dips that lead, trail, or fully overlap that of the planet. If the moon is leading/trailing, its transit dip starts before/after the ingress of the planet (see Fig.\,\ref{fig:moon_ring_transit}a), whereas if the moon is initially occulted by the planet during its ingress, then its transit causes a further dip within that of the planet. Detecting the transit signal of a moon ideally requires observing the transit of the entire hill sphere to cover all possible moon separation. Several tools (e.g., \texttt{LUNA} \cite{Kipping2011LUNA:Transits}, \texttt{pandora} \cite{Hippke2022Pandora:Algorithm}, \texttt{gefera} \cite{Gordon2022AnalyticStar})  have been developed to model the dynamic effects of moons in transit data. Other proposed methods to detect exomoons include: the  orbital sampling effect
of phase-folded light curves \citep{Heller2016MODELINGMOONS}, the Rossiter-McLaughlin effect \citep{Simon2010MethodsEffect}, direct imaging \citep{Cabrera2007DetectingEvents}, signatures of moon volcanic activity \citep{Ben-Jaffel2014TRANSITDIAGNOSIS}, and microlensing \citep{Bennett2014MOA-2011-BLG-262Lb:BULGE}.


\subparagraph{Detecting exorings}

The presence of rings around a planet can produce distinct imprints or anomalies in the transit lightcurve, such as shallow pre- and post-transit dips or asymmetric ingress and egress profiles. \citep{barnes2004, akin18}. Rings could even cause pre- and post-transit brightening due to forward-scattering \citep{barnes2004}. The exact scale and structure of the anomalies depend on the ring orientation, radial extent, opacity, particle size, and the presence of gap(s). Unlike moons, rings do not induce dynamical effects on the planetary transit, so transits can be phase-folded to increase the signal-to-noise ratio. If not viewed edge-on, a planet with rings will cause significantly deeper transits as the rings block out additional stellar light (see Fig.\,\ref{fig:moon_ring_transit}b). This can be misinterpreted as a planet with a larger radius, which  would result in an extremely low planetary density when combined with its mass measurement \citep{zuluaga}. However, a fit without rings will show ingress and egress structures in the residuals that can point to the possible presence of rings \citep[e.g.,][]{barnes2004, akin18}. Indeed, a growing number of planets with extremely low densities (called super-puffs) have been detected, and planetary rings offers one possible explanation for their inflated radii \citep{Piro2020}.

Modeling transits of ringed exoplanets requires high-precision and high-cadence photometry to extract the ring signatures. Several tools, such as \texttt{exoring} \cite{zuluaga}, \texttt{SOAP3.0} \cite{akin18}, \texttt{pyPplusS}, \cite{Rein2019FastRings}, \texttt{Pryngles}, and \cite{Zuluaga2022TheExorings} have been developed to model the effects of exorings in transiting planets. Other proposed methods to detect exorings aim to probe their signatures in scattered light \citep{Zuluaga2022TheExorings} and polarimetry \citep{Veenstra2025AExoplanets}.

\subsection{The search for exomoons and exorings}

Despite the strong expectations and decades of exomoon modeling, the observational confirmation of exomoons remains elusive. The major challenges with detecting exomoons and exorings include the infrequency of transits for long-period planets, which limits the number of observable transits to model the TTV/TDV patterns and transit configuration of an exomoon. This also limits the attainable photometric precision for probing exorings. However, numerous dedicated searches have laid important groundwork, refined detection techniques, and established upper limits on exomoon and exoring properties across a range of exoplanetary systems. These efforts have primarily focused on transit data, especially from the Kepler mission, which offered long, continuous, high-precision photometric baselines required for such analyses.

\medskip
The first systematic search for exomoons was the Hunt for Exomoons with Kepler (HEK) project \citep{Kipping2012TheProject}. This initiative applied Bayesian photodynamic modeling to selected Kepler planet candidates, searching for joint signatures of TTVs, TDVs, and light curve anomalies indicative of moons. The HEK analysis of hundreds of systems yielded non-detections or only marginal signals. Nonetheless, they placed valuable constraints on satellite-to-planet mass ratios and orbital configurations  \citep[e.g.,][]{Kipping2013ThePlanet, Kipping2013TheCandidates}. One of the prominent exomoon candidates from the project is the long-period giant planet Kepler-1625\,b \citep{Teachey2018HEK.I}. This planet was suggested to host a Neptune-sized companion \cite{Teachey2018EvidenceKepler-1625b} based on Kepler and Hubble Space Telescope (HST) observations that revealed a possible post-transit dip and a significant TTV. However, the interpretation remains controversial due to the limited data, potential systematics in the HST light curve, and alternative explanations \citep[e.g.,][]{Kreidberg2019NoSystem}. Further exomoon surveys \citep[e.g.,][]{Hippke2015ONPEAK,Kipping2022AnB-i} have examined hundreds of transiting planets and found no confirmed moons, setting upper bounds on the occurrence rate of Galilean-sized moons around exoplanets.  

\bigskip
A handful of studies have performed systematic and targeted searches to detect rings around exoplanets, identifying possible ring system candidates based on anomalous transit features in Kepler data \citep[e.g.,][]{heising,aizawa,Aizawa2018}. In many of the cases, it was possible to put valuable constraints on ring sizes and orientations. One notable case is J1407b, whose complex transit signal was proposed to be due to a substellar companion hosting a vast, multi-ring system with gaps possibly carved by exomoons \citep{ken}. However, no subsequent transits have been detected, and it is likely that the companion is not bound to the star. Another notable case is that of the long-period (542\,d) planet HIP-41378\,f which was targeted as a good candidate to probe for exorings due to its anomalously low density. Analysis of the Kepler K2 data revealed that its transits can be explained by a smaller planet with opaque rings extending out to 2.6 times the planetary radius \citep{Akinsanmi2020a}. Dynamical analysis of the target also found the ring scenario plausible \citep{Lu2024ThePlanet}. Further HST data revealed a flat spectrum that is consistent either with high-altitude hazes or rings around the planet \citep{Alam_et_al_2022}. It is still unclear whether the planet hosts rings, as there are other competing theories that can explain the inflated radii of super-puffs. The precision and wavelength span of JWST offer the potential to distinguish these scenarios. Other possible ring-hosting candidates include KIC 8462852 \citep[Boyajian's Star;][]{Ballesteros2017KIC2021} and PDS 110 \citep{Pinheiro2021}, although these systems are highly uncertain, and alternative explanations (e.g., dust or stellar variability) are often favored. A recent exoring survey by \citet{Umetani2025SearchTESS} estimates the occurrence rate of rings to be lower than 2\%.  

\medskip
Altogether the results from the exomoon and exoring searches (mostly in Kepler targets) suggest that large moons and extensive Saturn-like rings may not be very common around exoplanets, or our current detection techniques are not very sensitive to small or dynamically complex satellite systems. Moreover, these searches face limitations due to the rarity of long-period planet transits and the faintness of the targets. Distentangling the faint signals from other astrophysical noise sources also posed a considerable challenge. However, there are still avenues for the continued search for these satellite features: around brighter stars and the accumulation of more transits of long-period planets. These areas of improvement were recognized by the science team of the CHEOPS consortium. With its flexible pointing and high photometric precision, CHEOPS is uniquely positioned to follow long-period planets to probe for the presence of moons and rings. Therefore, an observational program  within the GTO was dedicated to identifying and following up on the best and targets to detect these elusive features. New long-period planets detected within the consortium (e.g., from the Mono- and Duotransit programs) are also probed for signatures of moons and rings. For instance, the recent confirmation of the 106\,d planet, TOI-2449\,b (Ulmer-Moll, under review), is being analyzed for anomalies that may reveal satellite features. More notable is the serendipitous confirmation of $\nu^2$ Lupi\,d, a 107\,d super-earth around the naked-eye star $\nu^2$\,Lupi hosting three planets \citep{Delrez2021}. The follow-up observation by CHEOPS covered the transit of a large fraction of the Hill sphere, which is as large as the Earth's, allowing to probe for the possible signal of an exomoon. The analysis of this single-epoch data ruled out a transiting moon larger than Mars \citep{Ehrenreich2023ACHEOPS}. This target is still being followed up by CHEOPS to gather more transit observations for a comprehensive moon search.



\section{Outlook}


Transiting planets on long orbits bring a wealth of information for the characterization of individual objects and the study of planet formation and evolution. The next generation of space-based missions 
will bring new detections of long-period transiting planets. The predicted yield for Neptune to Jupiter-sized planets with orbital periods larger than 50 days from the PLATO mission is a few hundreds of planets (e.g. \citealt{matuszewski_2023}). The Roman Space Telescope will deliver a few thousands of new transiting giant planets with semi-major axis larger than 0.3 au based on estimates from \citet{wilson_2023}. Additionally, Gaia DR4 will detect thousands of planets via astrometry, including gas giant planets located in the habitable zone of their host stars and beyond. Planets with inclinations close to 90 degrees could be detected in transit and estimates are of the order of a few tens of transiting planets \citep{perryman_2014a,sozzetti_2014}. Such detections would reveal unique objects : transiting temperate and cold gas giants which will be the closest known analogs of the solar system giant planets.%

Future research in giant exoplanet characterization will benefit greatly from improved evolution models and increasingly precise observational data. Evolution models are key to determining the bulk composition of giant exoplanets and linking atmospheric and interior properties. However, theoretical uncertainties remain, and refining these models to reflect the most realistic physical parameters is an ongoing challenge. Current and upcoming space missions like PLATO, JWST, ARIEL, and the Roman Space Telescope, along with ground-based facilities, are expected to play a transformative role. Precise planetary radii, stellar age determinations, and detailed atmospheric measurements will enable tighter constraints on bulk heavy-element content and clarify the relationship between atmospheric and interior compositions. Additionally, focused studies on giant planets with masses between about 0.3 and 2 Jupiter masses orbiting M-dwarf stars will be pivotal. These measurements will help distinguish different populations of giant planets and allow direct comparisons of bulk and atmospheric compositions across host star types. This will provide deeper insights into the processes of giant planet formation across a range of stellar environments.

While no exomoon or exoring has been confirmed with high statistical confidence, the concerted efforts have significantly advanced photodynamic modeling and light curve analysis techniques to improve detection sensitivity. Recent and upcoming space-based facilities
and next-generation ground-based observatories are expected to bring higher sensitivity, longer monitoring baselines, and/or multi-wavelength capabilities. Furthermore, direct imaging with instruments such as the Habitable Worlds Observatory \citep[HWO][]{} could potentially allow the detection of the thermal emission or reflected light from large exorings \citep{bowensrubin2025detectionexoringsreflectedlight}, and also large, possibly tidally heated, exomoons \citep{Limbach2013,Vanderburg_2018, Lazzoni2022}. These advances hold significant promise for the detection and characterization of these elusive features, enabling a deeper understanding of planet–moon co-evolution, satellite formation pathways, and the broader diversity of planetary systems.

\begin{acknowledgement}
We would like to thank Xinyi Song and Attila Simon for their management of the CHEOPS program on Moons and Rings and discussion that helped in writing this section. SM and BA acknowledge support from the Swiss National Science Foundation (SNSF) under grants \texttt{\detokenize{200020_215634}} and \texttt{\detokenize{PCEFP2_194576}}, respectively, and the National Centre for Competence in Research ‘PlanetS’ supported by SNSF under grants 51NF40\_182901 and
51NF40\_205606. This research used data from the NASA Exoplanet Archive, which is operated by the California Institute of Technology under contract with the National Aeronautics and Space Administration under the Exoplanet Exploration Program. Extensive use was also made of the Python packages \texttt{Jupyter} \citep{jupyter}, \texttt{Matplotlib} \citep{2007CSE.....9...90H}, and \texttt{NumPy} \citep{2020Natur.585..357H}.
\end{acknowledgement}



\small
\bibliography{references, akin_references, solene_planetS}
\end{document}